\documentclass[aps,preprint,superscriptaddress]{revtex4-2}

\usepackage{amsmath,amsfonts,amssymb}
\usepackage{enumitem}
\setlist[enumerate]{itemsep=0mm}

\usepackage[pdftex]{graphicx}
\usepackage{gensymb}
\usepackage{microtype}
\usepackage{bm}
\usepackage[normalem]{ulem}

\usepackage[svgnames]{xcolor}
\usepackage[
	colorlinks=True,linkcolor=DarkRed,citecolor=ForestGreen,urlcolor=MediumBlue,
	pdfstartview=FitH,bookmarks=False,pdfpagemode=UseNone
]{hyperref}

\newif\iffigure
\figuretrue

\newif\ifCMbool
\CMboolfalse

\ifCMbool
\newcommand{\CMs}[1]{\textcolor{red}{{\sout{#1}}}}
\else
\newcommand{\CMs}[1]{\textcolor{red}{{ }}}
\fi

\newcommand{\irte}{IrTe$_2$}

\begin{document}

\preprint{APS/123-QED}

\title{Ultrafast recovery dynamics of dimer stripes in IrTe$_2$}

\author{M. Rumo}
\affiliation{University of Fribourg and Fribourg Centre for Nanomaterials, Chemin du Mus\'ee 3, 1700 Fribourg, Switzerland}
\author{G.~Kremer}
\affiliation{University of Fribourg and Fribourg Centre for Nanomaterials, Chemin du Mus\'ee 3, 1700 Fribourg, Switzerland}
\affiliation{Universit\'e de Lorraine, CNRS, IJL, F-54000 Nancy, France}

\author{M. Heber}
\affiliation{Deutsches Elektronen-Synchrotron DESY, Notkestr. 85, 22607 Hamburg, Germany}
\affiliation{European XFEL, Holzkoppel 4, 22869 Schenefeld, Germany}
\author{N. Wind}
\affiliation{Deutsches Elektronen-Synchrotron DESY, Notkestr. 85, 22607 Hamburg, Germany}
\affiliation{Institut f\"ur Experimentelle und Angewandte Physik, Christian-Albrechts-Universit\"at zu Kiel, 24098 Kiel, Germany}

\author{C. W. Nicholson}
\affiliation{University of Fribourg and Fribourg Centre for Nanomaterials, Chemin du Mus\'ee 3, 1700 Fribourg, Switzerland}

\author{K. Y. Ma}
\affiliation{Max Planck Institute for Chemical Physics of Solids, N\"othnitzer Str. 40, 01187 Dresden, Germany}

\author{G. Brenner}
\affiliation{Deutsches Elektronen-Synchrotron DESY, Notkestr. 85, 22607 Hamburg, Germany}

\author{F. Pressacco}
\affiliation{Deutsches Elektronen-Synchrotron DESY, Notkestr. 85, 22607 Hamburg, Germany}

\author{M. Scholz}
\affiliation{Deutsches Elektronen-Synchrotron DESY, Notkestr. 85, 22607 Hamburg, Germany}

\author{K. Rossnagel}
\affiliation{Institut f\"ur Experimentelle und Angewandte Physik, Christian-Albrechts-Universit\"at zu Kiel, 24098 Kiel, Germany}
\affiliation{Ruprecht Haensel Laboratory, Deutsches Elektronen-Synchrotron DESY, 22607 Hamburg, Germany}

\author{F. O. von Rohr}
\affiliation{Department of Quantum Matter Physics, University of Geneva, 24 Quai Ernest-Ansermet, 1211 Geneva, Switzerland}

\author{D. Kutnyakhov}
\affiliation{Deutsches Elektronen-Synchrotron DESY, Notkestr. 85, 22607 Hamburg, Germany}

\author{C. Monney}
\email{claude.monney@unifr.ch}
\affiliation{University of Fribourg and Fribourg Centre for Nanomaterials, Chemin du Mus\'ee 3, 1700 Fribourg, Switzerland}

\date{\today}

\begin{abstract} 
The transition metal dichalcogenide \irte\ displays a remarkable series of first-order phase transitions below room temperature, involving lattice displacements as large as 20 percents of the initial bond length. This is nowadays understood as the result of strong electron-phonon coupling leading to the formation of local multicentre dimers that arrange themselves into one-dimensional stripes.
In this work, we study the out-of-equilibrium dynamics of these dimers and track the time evolution of their population following an infrared photoexcitation using free-electron lased-based time-resolved X-ray photoemission spectroscopy. First, we observe that the dissolution of dimers is driven by the transfer of energy from the electronic subsystem to the lattice subsystem, in agreement with previous studies.
Second, we observe a surprisingly fast relaxation of the dimer population on the timescale of a few picoseconds. By comparing our results to published ultrafast electron diffraction and angle-resolved photoemission spectroscopy data, we reveal that the long-range order needs tens of picoseconds to recover, while the local dimer distortion recovers on a short timescale of a few picoseconds. 
\end{abstract}

\maketitle

\section{Introduction}

Transition metal dichalcogenides represent an important class of materials in condensed matter physics with their layered quasi-two-dimensional structure, despite a simple chemical composition. Due to their large diversity, they have been continuously studied over the past 50 years, notably for the occurrence of structural phase transitions that are often the result of large electron-phonon coupling and/or an enhanced electronic susceptibility stemming from shallow $d$-bands \cite{Rossnagel2011}. In the case of strong electron-phonon coupling, a local-chemical-bonding picture is expected, for which the specific momentum dependence of the near-Fermi-level electronic structure does not play a dominant role \cite{McMillan1977}. In that case, the coherence length is short and the atomic displacements are large.

In the present paper, we focus on the case of \irte, a material that undergoes a series of first-order structural phase transitions below room temperature (RT).
Already at 280~K the crystal structure changes from a trigonal 1$T$ phase ($P$-3$m$1) to a monoclinic phase ($P$-1) involving a (5$\times$1$\times$5) superstructure \cite{Yang2012}. In the bulk this is followed by a second transition at 180~K into an (8$\times$1$\times$8) phase \cite{Ko2015}. Low-temperature (LT) measurements have revealed a (6$\times$1$\times$6) ground state \cite{Hsu2013, Takubo2018}, which can also be stabilized in Se doped \cite{Oh2013} and uniaxially strained crystals \cite{Nicholson2021}. Although the resistivity is reduced in the LT phases, \irte\ remains metallic at all temperatures. Each of the phases below 280~K exhibits a shortening of some of the Ir-Ir distances, giving rise to dimerized atoms arranged into one-dimensional stripes \cite{Pascut2014a}. The number of dimers per unit cell is characteristic to each of the LT phases and leads to one-dimensional stripes with a specific periodicity (see Fig.~\ref{fig_1}(a) and (b)) \cite{Ko2015}. Numerous angle-resolved photoemission spectroscopy (ARPES) studies have provided insights into the electronic structure of \irte\ \cite{Ootsuki2013, Ootsuki2017, Qian2014, Ko2015, Blake2015, Lee2017a, Rumo2020, Nicholson2021, Bao2021, Rumo2021, Mizokawa2022,Bao2021,Nicholson2021}. Extraction of the intrinsic electronic structure is complicated by the spatial coexistence of a ladder of nearly degenerate phases with periodicity $3n+2$ ($n=1, 2, 3,...$) at the surface with (5$\times$1), (8$\times$1) and (6$\times$1) the dominant phases in observations \cite{Hsu2013}. In addition, several X-ray photoemission spectroscopy (XPS) studies observed a clear impact of the dimer atomic distortion on the local environment of atoms, giving rise to satellite peaks appearing in Ir core-levels \cite{Qian2014, Ko2015, Rumo2020, Nicholson2021,Rumo2022}.
This naturally brings two questions: what is the mechanism stabilizing the single dimer formation and why do the dimers organize themselves in a particular stripe arrangement? 

The first question has been addressed already in depth in the last decade.
Initially proposed weak-coupling scenarios based on Fermi surface nesting \cite{Yang2012} appear to be incompatible with the dramatic 20 percent reduction in Ir bond lengths later observed in the LT phases \cite{Pascut2014a}.
A number of works have put forward ideas based on a local molecular-type picture involving either depolymerisation between the layers \cite{Oh2013}, or an effective Ir dimerization that produces bonding and anti-bonding 5$d$ states \cite{Pascut2014a, Pascut2014b} as part of a more complex multicentre bond \cite{Saleh2020}. Further support for this scenario was obtained by recent high-pressure diffraction measurements, where an observed modification of the Ir-Te angle provides conditions compatible with a ring-shaped bond across multiple sites \cite{Ritschel2022}. In both the LT and high-pressure calculations, the predicted bonding states appear distinctly split-off to higher binding energies from the density of states of the remaining undimerized Ir atoms, which was recently observed experimentally by Nicholson and coworkers \cite{Nicholson2024}.
Therefore, in the past few years, a consensual picture has been reached, for which local bonding-antibonding molecular-like orbitals involving both Te and Ir atoms are stabilized by a strong electron-phonon coupling \cite{Saleh2020,Ritschel2022,Nicholson2024}.

The second question is more delicate to answer. To the authors knowledge, theoretical studies about the structural arrangement of dimers in stripes of different periodicities have been relying on experimental data. As far as we know, there is no fundamental investigation on how different stripe patterns arise from isolated dimers, apart from a real-time scanning tunneling microscopy (STM) study of domain boundary evolutions near the phase transitions \cite{Mauerer2016}. Recent works performed on \irte\ monolayers on semiconducting \cite{Hwang2022a} and metallic substrates \cite{Asikainen2025} report that the dielectric and electronic environment of the substrate has a strong impact on the presence and periodicity of dimer stripes. Time-resolved studies using the pump-probe technique could potentially address this question, by suppressing dimers and their long-range order and tracking the intermediate states appearing during the relaxation back to the ground state. Ideta \textit{et al.} conducted a time-resolved electron diffraction study of the suppression and recovery of the superlattice peak related to the structural distortion, revealing relatively slow processes on the scale of tens of picoseconds \cite{Ideta2018a}.
Slow timescales were confirmed by a time-resolved ARPES study of Monney and coworkers, who concluded that the suppression of the ordered phase is driven by the flow of energy into the lattice following photoexcitation \cite{Monney2018}.
In addition to these experiments, Liu and coworkers performed time-dependent density functional theory (TDDFT) calculations to study the dimer dissociation mechanism following photoexcitation. Their work highlights the role of antibonding orbitals in this mechanism \cite{Liu2020,Liu2022}.

Here, we present a novel time-resolved perspective of the first low-temperature structurally distorted phase of \irte. Using time-resolved XPS, we monitor the ultrafast evolution of the population of dimers in the surface (5$\times$1) phase. This spectroscopic probe provides an atomic site-specific view on the dynamics of dimers, in contrast to previous studies \cite{Ideta2018, Monney2018}. The photoexcitation at 2.4 eV suppresses the dimers within 500 fs, in agreement with other time-resolved experiments sensitive to long-range order. However, the recovery of dimers occurs within a few picoseconds only. We conclude that the full relaxation to the ground state is a multi-step process, during which disordered dimers reappear first, before the restoration of long-range order.

\section{Experimental details}

Single crystals of \irte\ were grown using the self-ﬂux method \cite{Jobic1991, Fang2013} and were characterized by magnetic susceptibility and resistivity measurements reported in a previous publication \cite{Rumo2020}.
We performed time-resolved XPS experiments using a time-of-flight (ToF) momentum microscope (HEXTOF experimental station) at the plane grating monochromator beamline (PG2) \cite{Martins2006,Gerasimova2011} of the free electron laser FLASH, DESY, Hamburg \cite{Kutnyakhov2020}. To minimize pump-induced space charge effects, the momentum microscope was used in repeller mode \cite{kach2024}. Samples were cleaved in vacuum prior to the measurements and kept at a base temperature of 250 K. After correcting for the temporal pulse jitter from FLASH, the effective time resolution was 250 fs (full width at half maximum).
The beam spot size used to calculate the photoexcitation fluence was estimated using the real-space multiphoton photoemission signal measured by the momentum microscope on two different calibration samples (W(110) and Chessy sample (Plano GmbH))), including a correction for the higher-order process necessary for photoemission (see Ref. \cite{Dendzik2020} for more details). 
In this paper, values for the incident fluence are given. We estimated the systematic uncertainty to be about 20\%.

\iffigure
\begin{figure}
\begin{center}
\includegraphics[width=11.3cm]{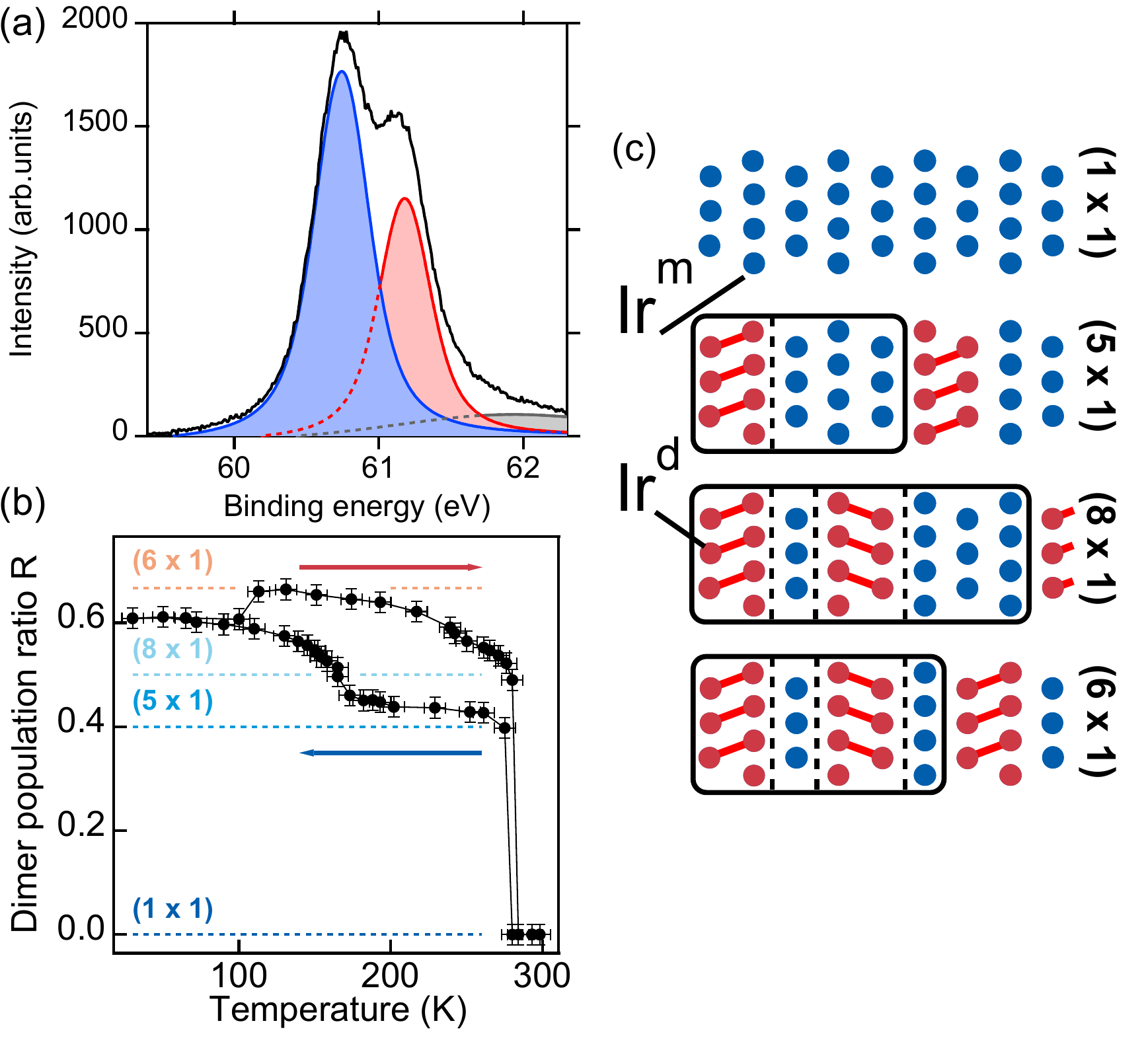}
\end{center}
\caption{\label{fig_1}
(a) Representative XPS spectrum of the Ir 4$f_{7/2}$ core-levels, taken at 250 K and 270 eV photon energy. The colored area are Voigt functions used to fit the spectrum and calculate the dimer population ratio $R$. The blue and red components are the Ir$^{m}$ monomer and Ir$^{d}$ dimer satellites, respectively. The grey component is used to fit a metallic asymmetric background. (b) Evolution of the dimer population ratio $R$ across the different first-order phase transitions upon cooling (blue arrow) and warming (red arrow) (data taken with 200 eV photon energy, graph adapted from Ref. \cite{Rumo2020}). These data were obtained in our previous study using samples from the same growth batch. The colored dashed horizontal lines represent the expected ratios $R$ for ideal surface phases, as depicted by the amount of Ir dimers (red balls) in graph (c) (adapted from Ref. \cite{Rumo2020}). Blue balls indicate monomer Ir atoms. 
}
\end{figure}
\fi

\section{Results}

Fig.~\ref{fig_1}(a) shows a typical Ir $4f_{7/2}$ core-level spectrum measured by XPS with a photon energy of 270 eV at 250~K at FLASH. At this temperature, \irte\ is in the surface (5$\times$1) phase and a prominent satellite peak (red shading) coming from Ir$^{d}$ dimerized states is observed at about 61.1 eV, in addition to the original peak (blue shading) of the Ir$^{m}$ monomer states. Atomic bonds in the dimers are reduced by up to 20\% in comparison to the RT phase, leading to a new local structural environment that induces chemical shifts in XPS. As shown in our previous publication \cite{Rumo2020}, the ratio $R$ = Ir$^d$/(Ir$^m$+Ir$^d)$ of the total XPS intensities under the dimer Ir$^d$ and monomer Ir$^m$ peaks is a spectroscopic marker of the relative population of dimers in each phase. Fig.~\ref{fig_1}(b) adapted from Ref. \cite{Rumo2020} displays the ratio $R$ upon cooling and heating across the several phase transitions of \irte. The real-space arrangement of dimers (in red) and monomers (in blue) is schematically shown in Fig.~\ref{fig_1}(c) for the different phases. Therefore, for the (5$\times$1) phase, a ratio $R=0.4$ is expected ideally, which is close to the experimental value of $R=0.38(2)$ in Fig.~\ref{fig_1}(a).


The main goal of the present study is to monitor the ultrafast dynamics of the dimer population in \irte\ using time-resolved XPS after a 515 nm photoexcitation (2.4 eV). The sample was kept at a base temperature of 250 K, in the surface (5$\times$1) phase as a ground state. A collection of time-resolved XPS spectra for a sample exposed to a photoexcitation fluence of 0.79 mJ/cm$^2$ is shown in Fig.~\ref{fig_2}(a). At zero time delay ($t_0$), laser-assisted photoemission (LAPE, a final-state artifact \cite{Miaja-Avila2006}) scatters photoelectrons to different energies, and notably at a 2.4 eV-higher kinetic energy. Later on, different photoexcitation effects appear. To make photoinduced effects more obvious, we plot in Fig.~\ref{fig_2}(b) the difference photoemission intensity map, i.e. the difference between XPS spectra taken at a given time delay minus an XPS spectrum averaged over all time delays before $t_0$. This reveals several interesting effects, in addition to the LAPE intensity change occurring at $t_0$.

\iffigure
\begin{figure*}
\begin{center}
\includegraphics[width=16cm]{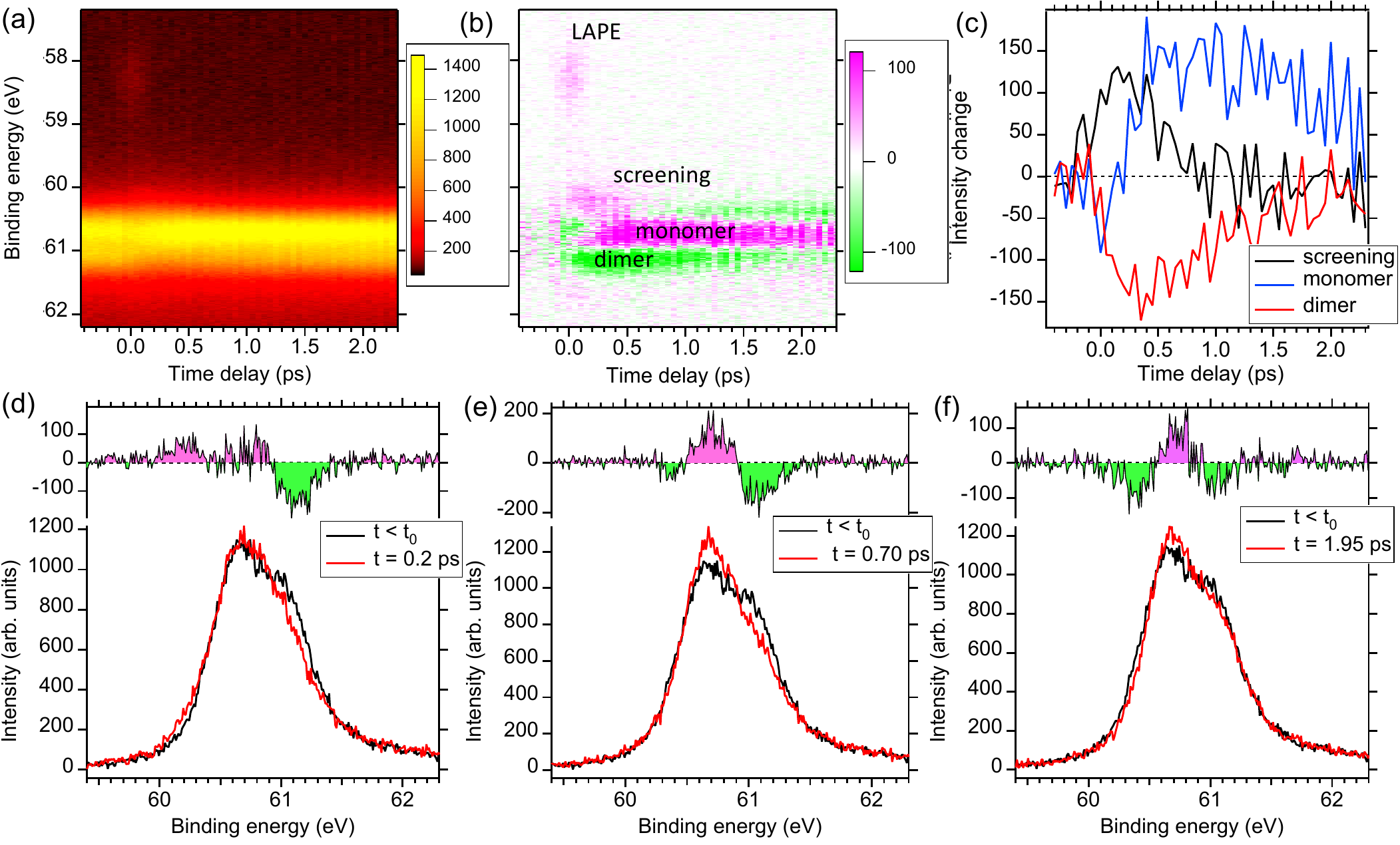}
\end{center}
\caption{\label{fig_2}
(a) Time-resolved XPS data of the Ir 4$f_{7/2}$ core-levels displayed as a function of pump-probe time delay relative to $t_0$. All data on this figure have been taken at 250 K, with a photon energy of 270 eV and a photoexcitation fluence of 0.79 mJ/cm$^2$. (b) Difference image plot showing data after subtraction of an average of data before $t_0$. This emphasizes 4 different transient effects described in the main text. (c) Transient intensity curves integrated on the data of graph (a) over $\pm 100$ meV around 60.15 eV for screening, 60.70 eV for monomer and 61.15 eV for dimer intensities.
(d),(e),(f) XPS spectra at specific time delays (red curves), in comparison with the spectrum averaged before $t_0$ (black curve). The differences between the two spectra are shown on top. 
}
\end{figure*}
\fi

The intensity decrease at 61.1 eV binding energy relates to the dimer core-level intensity changes. This indicates a transient suppression of dimer states in \irte. It is concomitant with an intensity increase at 60.7 eV that conversely indicates a transient monomer increase. Selected representative XPS spectra are shown in Fig.~\ref{fig_2} (d), (e), (f). One can clearly see the relative intensity change of the dimer vs. monomer peaks at 0.70 ps and 1.95 ps, going up to 15\%. However, the XPS spectrum at 0.2 ps has a different lineshape, with an asymmetry at lower binding energy. Such an effect has already been observed in XPS following photoexcitation in WSe$_2$ \cite{Dendzik2020}. The extra charge carriers generated around the chemical potential are enhancing the screening of the Coulomb interaction and leading to a lineshape asymmetry. The duration of this lineshape asymmetry correlates  with the population of excited electrons above the chemical potential. From this asymmetric signal appearing in the raw data, we therefore infer an excited carrier population lasting for about 1 ps, see black curve in Fig.~\ref{fig_2}c. We also plot on the same graph the transient evolution of the dimer and monomer related signal (this is obtained here as XPS signal integrated within a 200-meV wide interval around the respective peak positions). Interestingly, we observe at $t_0$ a rapid increase of the peak asymmetry due to free charge carrier population above the Fermi level. However, the changes in the dimer and monomer peak intensity are delayed with respect to the charge carrier increase. This is a first indication that the mechanism responsible for the dimer suppression is not directly related to the transient charge carrier increase. The relaxation time of the monomer and dimer population seems to be in the range of 1-3 ps at this fluence. A finer analysis of the dimer and monomer signals is given below using peak fitting.


We now turn to a systematic fluence-dependent study of the transient dimer suppression as observed in time-resolved XPS. We have gathered six datasets like the one presented in Fig.~\ref{fig_2}(a) for six different fluences ranging from 0.30 to 0.84 mJ/cm$^2$. For this purpose, we fit each XPS spectrum with 3 Voigt functions, as shown in Fig.~\ref{fig_1}(a). We plot in Fig.~\ref{fig_3}(a) and (b) the resulting dimer population ratio $R$ as a function of time delay. This collection of graphs represents the main experimental result of our study. For the lowest fluence of 0.30 mJ/cm$^2$, the ratio $R$ does not change within the sensitivity of our experiment. For higher fluences, we observe that the dimer population ratio $R$ decreases from its nominal value (dashed line) to a minimal value $R_\text{min}$ and then recovers back to its nominal value with a certain relaxation time mainly contained in the time delay window plotted here.
Moreover, the minimal value of the dimer ratio $R_\text{min}$ decreases with increasing fluence. More strikingly, we see that the necessary time to reach $R_\text{min}$ increases with the fluence (except for the dataset at 0.70 mJ/cm$^2$, which might be due to an improper pump-probe spatial overlap). This somewhat unusual observation for the ultrafast suppression of an ordered phase has already been explained in a time-resolved ARPES study of \irte, using a two-temperature model \cite{Monney2018}. This demonstrates that the suppression of the dimer population is driven by the photoexcitation energy flowing from the electronic subsystem into the lattice one. In other words, it is driven by the transient lattice temperature, which takes a longer time to reach its maximum at larger fluences.

\iffigure
\begin{figure}
\begin{center}
\includegraphics[width=13cm]{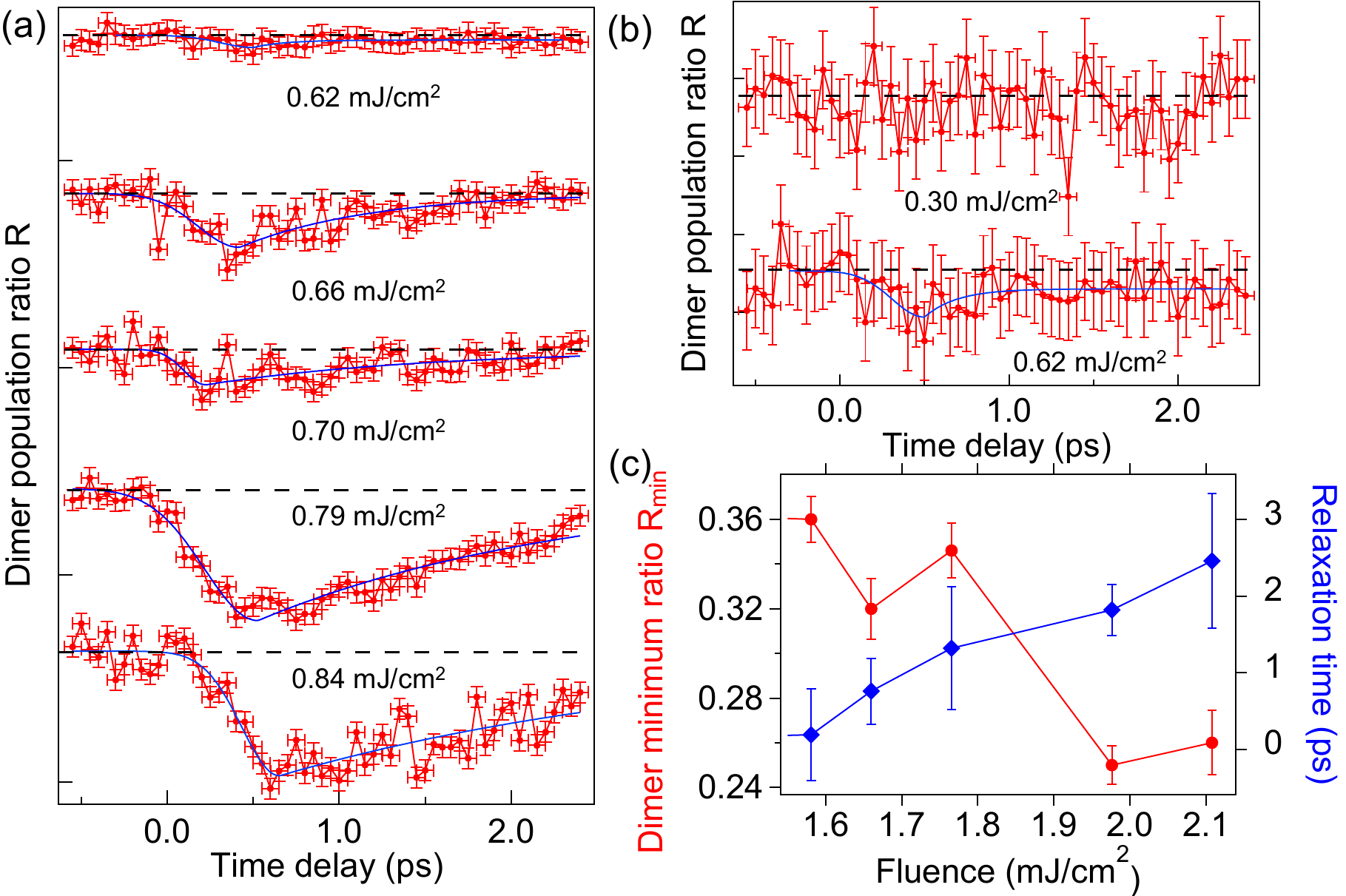}
\end{center}
\caption{\label{fig_3}
(a) Transient dimer population ratio $R$ for different photoexcitation fluences, together with their fit (blue). Curves are offset by multiples of 0.15 for clarity. The tick spacings on the vertical axis correspond to 0.2. (b)  Same for the lower two photoexcitation fluences. Curves are offset by multiples of 0.04 for clarity. The tick spacings on the vertical axis correspond to 0.02. (c) Minimum ratio $R_\text{min}$ and relaxation time extracted from the fits in graph (a) as a function of fluence.
}
\end{figure}
\fi

We performed a more quantitative analysis by fitting the transient dimer populations $R$ with an error function for the onset of the suppression and a single-exponential for the relaxation (see the blue curves in Fig.~\ref{fig_3}(a),(b)). We plot in Fig.~\ref{fig_3}(c) the obtained relaxation time and dimer minimal population $R_\text{min}$ as a function of fluence. The dimer minimal population decreases with increasing fluence, while the relaxation time increases up to 2.5 ps. This relaxation time is actually surprisingly short for a maximum fluence of nearly 1 mJ/cm$^2$. At the same fluence, the dimer population is substantially suppressed, reaching 63\% of its nominal value after about 600 fs.

\iffigure
\begin{figure*}
\begin{center}
\includegraphics[width=17cm]{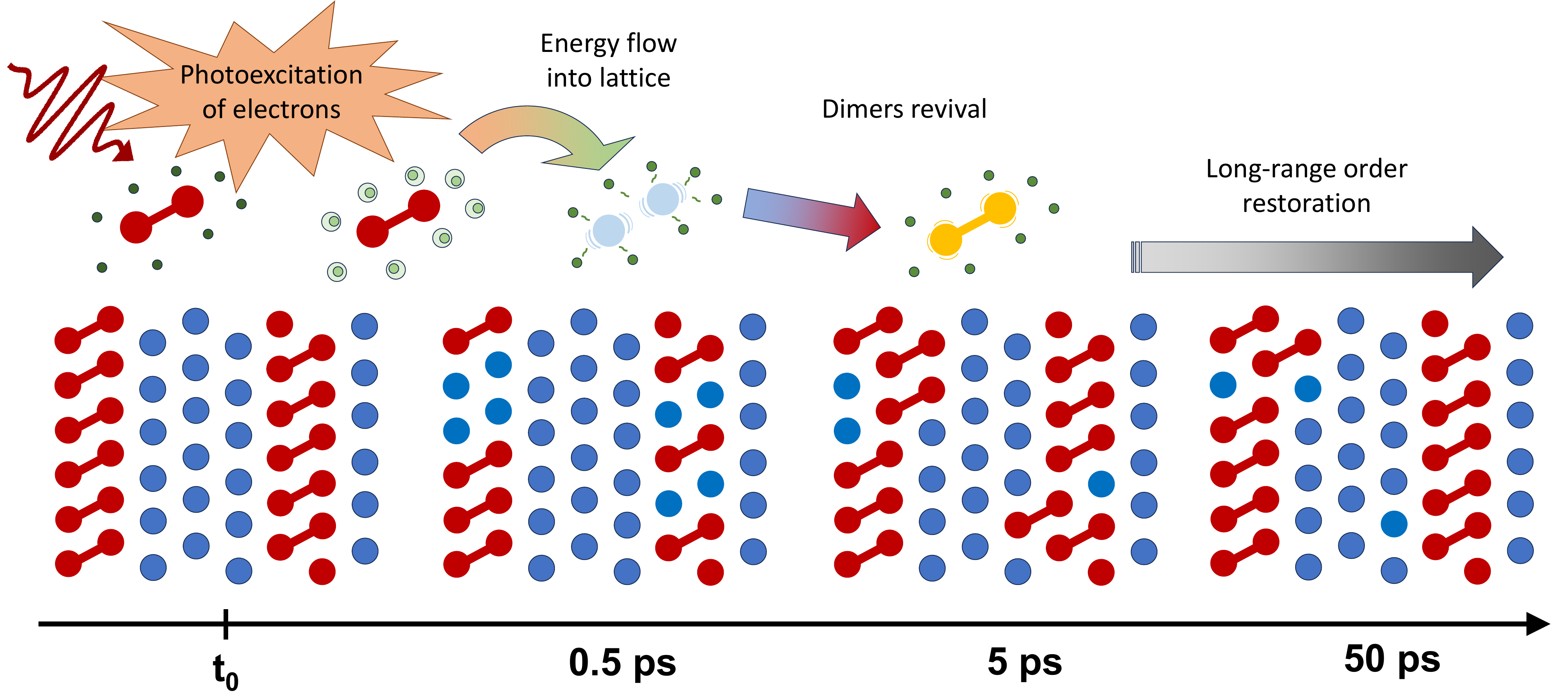}
\end{center}
\caption{\label{fig_4}
Cartoon describing the ultrafast dynamics of the structural phase transition in \irte\ and the recovery of the long-range order of dimers after a few-eV photoexcitation.} 
\end{figure*}
\fi

\section{Discussion}

It is insightful to compare our results with other probes of the non-equilibrium dynamics of the structurally distorted phases in \irte. First, we recall results from a previous time-resolved ARPES study of the low-energy electronic structure of \irte\ \cite{Monney2018}. In this study, the ultrafast dynamics of a specific ARPES feature related to the low-temperature phases was tracked, together with the transient electronic temperature. It was shown that the suppression of the structurally distorted phase is dictated by the transient lattice temperature, and not directly by the accumulation of excited charge carriers above the Fermi level. The relaxation time of these excited carriers was about 500 fs, in agreement with the carrier-induced XPS peak asymmetry of Fig.~\ref{fig_2}c. Most interestingly, the relaxation time of the ARPES signal related to the structurally distorted phase was clearly much longer than any timescale observed in our time-resolved XPS data at similar fluences.
While XPS probes the population of dimers, independently on whether they are ordered or not, ARPES is sensitive to the wave function of an electronic state originating from the ordered dimers. The different timescales for the relaxation of the (5$\times$1) phase observed in XPS vs. ARPES might then be related to the necessary time for dimers to re-establish a long-range order.

To test this hypothesis, we now recall the results of the time-resolved electron diffraction experiment by Ideta and coworkers performed on samples as thin as 20 nm \cite{Ideta2018a}. They followed the transient evolution of the superlattice peak associated with the (5$\times$1$\times$5) reconstruction at 200~K after a partial suppression of the stripe order. They observed a substantial decrease of the intensity of the superlattice peak within about 300 fs, and a partial recovery taking place over about 7 ps. However, the fully ordered state did not recover in less than 20 ps, still exhibiting a 50\% suppression at longer time delays. The fluence used in this time-resolved diffraction experiment was rather low (0.55 mJ/cm$^2$). This is slightly smaller than the minimum fluence at which we observe changes in XPS. It allows us to reject the possibility that the long-lasting effect observed in Ideta and coworkers' experiment is due to using a much larger photoexcitation fluence than in our experiment. Again, the very different timescales for relaxation between electron diffraction and XPS points to their different sensitivity to long-range order and dimer population, resp..

The scenario arising from our time-resolved XPS experiment, and the above comparison to results from other time-resolved probes, is displayed in the schematic of Fig.~\ref{fig_4}. At $t_0$, the 2.4 eV photons of the pump pulse populate a large spectrum of electronic orbitals in \irte\ above the Fermi level, and including the antibonding orbitals of the dimer states. While the transient occupation of these antibonding orbitals is likely to preferentially destabilize the dimers more efficiently \cite{Ideta2018a}, our present study, together with the time-resolved ARPES study of Monney and coworkers \cite{Monney2018}, clearly demonstrates that the main actor of the dimer suppression using a 2.4-eV photoexcitation is the lattice transiently overheated by the transfer of energy from the electronic subsystem. Our local probe of the dimer population indeed evidences a delayed onset of the related XPS peak response with respect to $t_0$ (Fig.~\ref{fig_3}). Thus, suppression of dimers occurs on a timescale of about 0.5 ps. The dimers reappear within less than 5 ps, likely due to heat dissipation into the rest of the solid. However, the stripe order has not fully recovered on this timescale, and the dimers have adopted a rather disordered state. This explains why ARPES and electron diffraction do not see a fully recovered phase. While the creation of a local dimer leads to a large gain of energy, the short-range ordered state is not energetically the most favorable state \cite{Pascut2014,Saleh2020}. The long-range stripe order of dimers is recovering on a much longer timescale. The ultrafast electron diffraction experiment by Ideta and coworkers indicates that it does not occur before 50 ps, even for a modest photoexcitation fluence \cite{Ideta2018a}. The kinematics of the dimer reorganization towards the stripe order is a complicated process, involving the propagation of domain boundaries, as evidenced by a real-time STM study \cite{Mauerer2016}. The extension of such a study to the pump-probe regime and 100 ps timescale would be very instructive.

In their TDDFT simulations, Liu and coworkers observe that the fastest way to suppress the dimer is achieved by artificially placing electrons in the antibonding orbitals above the Fermi level of \irte\ \cite{Liu2022}. The resulting coherent lattice motion dissociates the dimer in 300 fs. However, they also simulate the photoexcitation by a 400-nm laser pulse that partially dissociates dimers in about 2 ps, without any coherent lattice motion. They highlight an interplay between phonons and electrons scattered into antibonding orbitals leading to a slow self-amplification process. This second scenario is more consistent with our experimental observations. The work of Liu and coworkers suggests repeating time-resolved experiments with a narrow-bandwidth optical excitation tuned resonantly to  antibonding orbitals to achieve a faster phase transition.

%
%

\section{Conclusion}

In this work, we have studied the ultrafast atomic-site specific dynamics of a first-order structural phase transition in \irte\ with time-resolved XPS. We observe a partial suppression of dimers after a non-resonant photoexcitation. The timescales of both their suppression and their recovery are increasing with photoexcitation fluence. The recovery dynamics of the dimer population measured with XPS is one order of magnitude faster than the recovery probed by electron diffraction. We conclude that the full recovery of the long-range stripe order of dimers in \irte\ following photoexcitation is a multistep process involving a fast recovery of partially disordered dimers that reorder slowly on a larger spatial scale.
Our work emphasizes the added value of combining different techniques to probe the out-of-equilibrium dynamics of ordered phases. We hope that our results will stimulate other time-resolved studies of the dimer dynamics in \irte\ and anticipate that time-resolved STM would bridge existing observations made with time-resolved XPS and ultrafast electron diffraction.

\section{Acknowledgments}
We acknowledge DESY (Hamburg, Germany) for the provision of experimental facilities. Parts of this research were carried out at FLASH using the PG2 beamline. D.K., M.H., N.W., K.R. acknowledges support via the Collaborative Research Center (CRC) 925, Project ID 170620586 (Project B2). We thank Holger Meyer and Sven Gieschen for support with the instrumentation. C.M., G.K., M.R., C.W.N. acknowledges support by the SNSF grant No. $PZ00P2\_ 154867$. 

\bibliography{library}

\end{document}